\documentclass[preprint,aps,pra,showpacs,floatfix]{revtex4-1}
\usepackage{amsmath}
\usepackage{times}
\usepackage {euscript}
\usepackage{graphicx}
\usepackage{amsthm}
\usepackage{epsfig}
\usepackage{bm}
\usepackage{multirow}
\usepackage{longtable}
\newcommand{\levels}{
\left[\left(1s 2s\right)_{0}n\kappa\right]_{1/2}
}
\newcommand{\levelp}{
\left[\left(1s 2p_{1/2}\right)_{0}n\kappa\right]_{1/2}
}
\newcommand{\chii}{
\chi_{1/2\mu_{i}}\left(\boldsymbol{\nu}\right)
}
\newcommand{\chif}{
\chi_{1/2\mu_{f}}\left(\bf n\right)
}
\graphicspath{{figures/}}
\begin{document}
\thispagestyle{empty}
\title{
Parity nonconservation effect in the resonance elastic electron 
scattering on heavy He-like ions
}
\author{V.~A.~Zaytsev$^{1,2}$,
        A.~V.~Maiorova$^{1}$,
        V.~M.~Shabaev$^{1}$,
        S.~Tashenov$^{3}$,
        and Th.~St\"ohlker$^{4,5,6}$}
\affiliation{
$^1$ Department of Physics, St. Petersburg State University,
Ulianovskaya 1, Petrodvorets, 198504 St. Petersburg, Russia \\
$^2$ SSC RF ITEP of NRC 
\textquotedblleft Kurchatov Institute\textquotedblright,
Bolshaya Cheremushkinskaya 25, Moscow, 117218, Russia \\
$^3$ Physikalisches Institut, Universit\"at Heidelberg, D-69120 
Heidelberg, Germany\\
$^4$ GSI Helmholtzzentrum f\"ur Schwerionenforschung GmbH, D-64291 
Darmstadt , Germany\\
$^5$ Helmholtz-Institut Jena, D-07743 Jena, Germany\\
$^6$ Institut f\"ur Optik und Quantenelektronik, 
Friedrich-Schiller-Universit\"at Jena, D-07743 Jena, Germany
\vspace{10mm}
}
%
\begin{abstract}
We investigate the parity nonconservation effect in the elastic 
scattering of polarized electrons on heavy He-like ions, being initially 
in the ground state.
The enhancement of the parity violation is achieved by tuning the energy
of the incident electron in resonance with quasidegenerate 
doubly-excited states of the corresponding Li-like ion.
We consider two possible scenarios.
In the first one we assume that the polarization of the scattered 
electron is measured, while in the second one it is not detected.
\end{abstract}
%
\pacs{34.80.Lx, 11.30.Er}
\maketitle
\section{INTRODUCTION}
\label{sec:INT}
Investigations of the parity violation in the domain of atomic 
physics originate from consideration of the PNC effects in neutral 
systems (see Refs.~\cite{Khriplovich, Khriplovich_PST112_52:2004, 
Ginges_PR397_63:2004} and references therein).
The most accurate up to date measurement of the PNC was achieved for 
$^{133}$Cs atom~\cite{Wood_S275_1759:1997, Bennett_PRL82_2484:1999}.
These experimental data being coupled with the corresponding theoretical 
calculations of the same accuracy level (see 
Refs.~\cite{Shabaev_PRL94_213002:2005, Porsev_PRL102_181601:2009, 
Dzuba_PRL109_203003:2012} and references therein) provided the best 
verification of the electroweak sector of the Standard Model at 
low-energy regime.
However, the precise calculations of the PNC effects in neutral systems 
are very difficult task.
For this reason, the investigations of the PNC effects in heavy 
few-electron systems where the interelectronic interaction can be 
calculated accurately by means of the perturbation theory in the 
parameter $1/Z$ ($Z$ is the nuclear charge number) seem very promising.
%
\\
\indent 
%
Gorshkov and Labzowsky~\cite{Gorshkov_JL19_394:1974} were first who 
considered highly-charged ions as a proper tool for measuring the PNC
effect.
To date, various theoretical scenarios were proposed to study the P-odd 
asymmetry in highly-charged ions.
The PNC effect in the process of Auger decay of the He-like uranium was 
studied by Pindzola~\cite{Pindzola_PRA47_4856:1993}.
Gribakin \textit{et al.}~\cite{Gribakin_PRA72_032109:2005} discussed the
parity violation in the process of dielectronic recombination of 
polarized electrons with H-like ions.
A similar process for the case of He-like ions was investigated in 
Ref.~\cite{Zaytsev_PRA89_032703:2014}.
The PNC effect in the process of radiative recombination of electron 
with H-like ions was studied in several
works~\cite{Maiorova_JPB42_205002:2009, Maiorova_JPB44_225003:2011, 
Gunst_PRA87_032714:2013}.
The parity violation on the laser-induced transition was considered 
for heavy He-like ions in Ref.~\cite{Shabaev_PRA81_052102:2010}
and for heavy Be-like ions in Ref.~\cite{Surzhykov_PST156_014027:2013}.
%
%
\\
\indent
%
Though the PNC effect in highly-charged ions was extensively studied, 
the influence of the weak interaction on the process of electron 
scattering by a heavy ion has not yet been investigated.
In the present work we study the PNC effect in the elastic scattering of
polarized electrons by heavy He-like ions, being initially in the ground 
state.
In order to enhance the parity violation we assume that the energy of 
the incident electron is tuned in resonance with close-lying
opposite-parity $\levels$ and $\levelp$ states of the corresponding 
Li-like ions~\cite{Zaytsev_PST156_014028:2013}.
%
\\
\indent
%
The relativistic units ($m_{\rm e} = \hbar = c = 1$) and the Heaviside 
charge unit ($\alpha = e^{2}/\left(4\pi\right)$) are used in the paper.

\section{BASIC FORMALISM}
\label{sec:BF}
\begin{figure}[h!]
\caption{
Geometry for the resonance elastic electron scattering in the ion rest 
frame.
The reaction plane is formed by ${\bf p}_{i}$ and ${\bf p}_{f}$ vectors,
which denote the momentums of the incident and outgoing electrons, 
respectively.
The normal to this plane is described by the unit vector 
$\boldsymbol{\eta} = \left[\boldsymbol{\nu} 
\times {\bf n} \right] / \left|\left[\boldsymbol{\nu} \times {\bf n} 
\right]\right|$ where $\boldsymbol{\nu} = {\bf p}_{i} / 
\left|{\bf p}_{i}\right|$ and ${\bf n} = {\bf p}_{f} / 
\left|{\bf p}_{f}\right|$.
}
\includegraphics[trim=0 0 0 0, clip, width = 1.0\textwidth]
{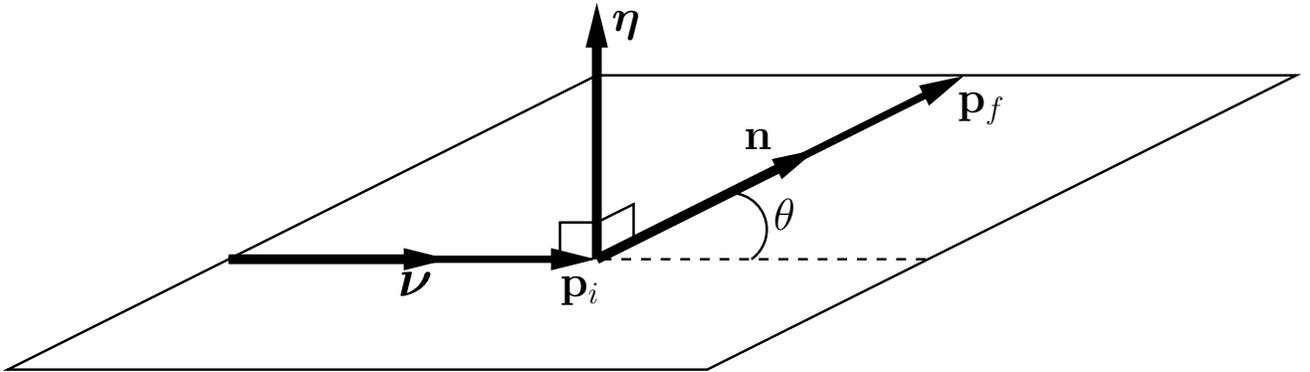} 
\label{fig:geometry}
\end{figure}

%
We consider the resonance elastic scattering of an electron with 
asymptotic four-momentum $\left(\varepsilon, {\bf p}_{i}\right)$ 
and polarization $\mu_{i}$ by a heavy He-like ion being initially in the 
ground $\left(1s\right)^{2}$ state.
It is assumed that the electron energy is tuned in resonance with 
doubly-excited quasidegenerate opposite-parity $d_{1}$ 
or $d_{2}$ states.
The scattered electron is characterized by four-momentum 
$\left(\varepsilon, {\bf p}_{f}\right)$ and polarization 
$\mu_{f}$.
%
\\
\indent
%
%
Let us start with the consideration of the parity conserving part of the
process amplitude.
We construct this amplitude by means of the $1/Z$ perturbation theory up 
to the second order:
\begin{equation}
\tau_{\mu_{f} \mu_{i}}^{\rm PC} = \tau^{(0)}_{\mu_{f} \mu_{i}} 
                                + \tau^{(1,\ \rm dir)}_{\mu_{f} \mu_{i}}
                                + \tau^{(1,\ \rm exc)}_{\mu_{f} \mu_{i}}
                                + \tau^{(2)}_{\mu_{f} \mu_{i}},
\end{equation}
where the first order contribution is separated into two terms which 
correspond to the direct and exchange parts of the interelectronic
interaction.
The sum of the zero-order and direct first-order terms can be 
written as follows~\cite{Akhiezer:1965}:
\begin{equation}
  \tau^{(0)}_{\mu_{f} \mu_{i}} 
+ \tau^{(1,\ \rm dir)}_{\mu_{f} \mu_{i}} = 
\chi^{\dagger}_{1/2\mu_{f}}\left(\bf n\right)
\left(
A + 2B \boldsymbol{\eta} \cdot \mathbf{S}
\right)
\chi_{1/2\mu_{i}}\left(\boldsymbol{\nu}\right),
\end{equation}
where $\bf S$ is the spin operator, $\boldsymbol{\nu}$ and $\bf n$ 
are the unit vectors in the ${\bf p}_{i}$ and ${\bf p}_{f}$ directions
(see Fig.~\ref{fig:geometry}), respectively, and 
$\boldsymbol{\eta} = \left[\boldsymbol{\nu} \times {\bf n} 
\right] / \left|\left[\boldsymbol{\nu} \times {\bf n} \right]\right|$.
The two-component $\chii$ function is an eigenfunction of the 
${\bf S} \cdot \boldsymbol{\nu}$ operator with an eigenvalue 
$\mu_{i}$ and $\chif$ satisfies 
$\left({\bf S} \cdot {\bf n} \right) \chif = \mu_{f} \chif$.
The scattering amplitudes $A$ and $B$ are defined 
as~\cite{Berestetsky:2006}:
\begin{eqnarray}
A & = & \frac{1}{2 {\rm i} p_{f}} 
\sum_{l = 0}^{\infty}
\left\lbrace
\left(l + 1\right)\left[\exp(2{\rm i}\delta_{l+1/2,l}) - 1 \right]
+ l\left[\exp(2{\rm i}\delta_{l-1/2,l}) - 1 \right]
\right\rbrace P_{l}\left(\cos\theta\right),
\label{eq:amplitude_A}
\\
B & = & \frac{1}{2 p_{f}} 
\sum_{l = 1}^{\infty}
\left[
\exp(2{\rm i}\delta_{l+1/2,l}) - \exp(2{\rm i}\delta_{l-1/2,l})
\right]
P_{l}^{1}\left(\cos\theta\right).
\label{eq:amplitude_B}
\end{eqnarray}
Here $p_{f}$ is the momentum of the scattering 
electron, $P_{l}$ and $P_{l}^{1}$ are the Legendre polynomials and 
associate Legendre functions, respectively, and $\theta$ is the 
scattering angle.
The phase shifts $\delta_{j,l}$ for the total angular $j$ and the 
orbital $l$ momenta are determined from the asymptotic behaviour of the 
Dirac equation solutions in the scattering potential 
$V(r) = V_{\rm nuc}(r) + V_{\rm scr}(r)$.
Here $V_{\rm nuc}$ is the electrostatic potential of the extended 
nucleus and $V_{\rm scr}$ is the screening potential of the $(1s)^{2}$ 
shell:
\begin{equation}
V_{\rm scr}(r) = 2 \alpha \int_{0}^{\infty} 
\frac{dr'}{r_{>}}
\left[ G_{1s}^{2}(r') + F_{1s}^{2}(r') \right],
\end{equation}
where $r_{>}$ is the greater of $r$ and $r'$, $G_{1s}(r)$ and 
$F_{1s}(r)$ are the upper and lower components of the radial wave 
function of one-electron $1s$ state, respectively.
Since $V(r) \sim \left(Z - 2\right) / r$ for large $r$, the scattering 
amplitudes defining by Eqs.~(\ref{eq:amplitude_A}) 
and~(\ref{eq:amplitude_B}) are divergent as they stand.
Nevertheless, one can obtain the convergent expression for $A$ and $B$
utilizing the regularization procedure~\cite{Mott_PRS135_429:1932, 
Dogget_PR103_1597:1956, Sherman_PR103_1601:1956, 
Gluckstern_JMP5_1594:1964} which deals with the pure Coulomb potential.
The deviation of the scattering potential from the Coulomb one is 
accounted for using the method described in 
Ref.~\cite{Maiorova_JPB43_205006:2010}.
%
%
\\
\indent
%
%
The exchange first-order amplitude $\tau_{\mu_{f}\mu_{i}}^{(1,\ \rm exc)}$ 
is constructed by subtraction of the terms corresponding to the direct 
part of the interelectronic interaction from 
$\tau_{\mu_{f}\mu_{i}}^{(1)} = (2\pi)^{2} \varepsilon \left\langle 
\Psi_{f} \left| I \right| \Psi_{i} \right\rangle$ (see 
Refs.~\cite{Shabaev_PRA50_4521:1994, Shabaev_PR356_119:2002} for details).
Here $I$ is the operator of the interelectronic interaction, 
$\left| \Psi_{i} \right\rangle$ and $\left| \Psi_{f} \right\rangle$ are
the wave functions of the initial and final states of the system, 
respectively.
Due to the fact that for heavy highly-charged ions the electron-electron
interaction is suppressed by a factor $1/Z$ compared to the 
electron-nucleus Coulomb interaction, we can utilize the one-electron 
approximation.
In this approach the wave functions of the initial and final states are 
given by
\begin{eqnarray}
\Psi_{{\bf p}\mu, JM} \left(
{\bf x}_{1},{\bf x}_{2},{\bf x}_{3} \right)
&=& A_{N} \sum_{\mathcal{P}} (-1)^{\mathcal{P}} \mathcal{P} 
\sum_{m_{1}m_{2}}
C_{j_{1}m_{1},\ j_{2}m_{2}}^{JM}
\psi_{n_{1}\kappa_{1}m_{1}}\left({\bf x}_{1}\right)
\psi_{n_{2}\kappa_{2}m_{2}}\left({\bf x}_{2}\right)
\psi_{{\bf p} \mu}\left({\bf x}_{3}\right).
\label{eq:wf_3e_1c2d}
\end{eqnarray}
Here $\psi_{n\kappa m}$ is the one-electron bound-state 
Dirac wave function and $\psi_{{\bf p}\mu}$ is the continuum Dirac state 
wave function with asymptotic momentum $\bf p$ and helicity $\mu$ (spin 
projection onto the momentum direction).
The normalization factor $A_{N} = 1/\sqrt{2\cdot 3!}$ for equivalent bound 
electrons and $A_{N} = 1/\sqrt{3!}$ otherwise, $C_{j_{1}m_{1},\ 
j_{2}m_{2}}^{JM}$ is the Clebsch-Gordan coefficient, 
$(-1)^{\mathcal{P}}$ is the permutation parity, and $\mathcal{P}$ is the 
permutation operator.
The explicit expression for the continuum Dirac wave function can be 
written as~\cite{Rose:1957, Eichler_PR439_1:2007}
\begin{equation}
\psi^{(\pm)}_{{\bf p}\mu}\left({\bf r}\right) = 
\frac{1}{\sqrt{4\pi}} \cdot
\frac{1}{\sqrt{\varepsilon {\rm p}}} 
\sum_{\kappa m_{j}} 
C^{j\mu}_{l0,\ 1/2\mu} 
{\rm i}^{l} 
\sqrt{2l + 1}
e^{\pm{\rm i}\delta_{j,l}}
D_{m_{j}\mu}^{j}({\bf z} \rightarrow {\bf p})
\psi_{\varepsilon\kappa m_{j}} \left({\bf r}\right),
\end{equation}
where the upper (lower) sign corresponds to the incoming (outgoing) 
electron and $\kappa = (-1)^{j+l+1/2}(j+1/2)$ is the Dirac quantum 
number.
The Wigner matrix $D_{MM'}^{J}({\bf z} \rightarrow {\bf p})$ 
(see Refs.~\cite{Rose:1957, Varshalovich:1988} for details) rotates
the $\bf z$ axis into the $\bf p$ direction.
%
%
\\
\indent
%
%
The second-order amplitude, corresponding to the dielectronic 
recombination into one of doubly excited $d_{1}$ or $d_{2}$ states with 
subsequent Auger decay, is given by~\cite{Shabaev_PRA50_4521:1994,
Shabaev_PR356_119:2002}
\begin{equation}
\tau_{\mu_{f}\mu_{i}}^{(2)} = (2\pi)^{2} \varepsilon \sum_{k = 1,2}
\sum_{M_{d_{k}}} \frac{
\left\langle \Psi_{f} \left|I\right| \Psi_{d_{k}} \right\rangle
\left\langle \Psi_{d_{k}} \left|I\right| \Psi_{i} \right\rangle}
{E_{i} - E_{d_{k}} + {\rm i}\Gamma_{d_{k}}/2},
\label{eq:amplitude_2}
\end{equation}
where $E_{d_{k}}$ is the energy of the $d_{k}$ state, $E_{i} = 
E_{(1s)^{2}} + \varepsilon$ is the energy of the initial state, 
$\Gamma_{d_{k}}$ is the total width and $M_{d_{k}}$ is the momentum 
projection of the $d_{k}$ state.
The wave functions of the $d_{1}$ and $d_{2}$ states in the one-electron 
approximation are given by
\begin{eqnarray}
\nonumber \Psi_{J(J')M} \left(
\mathbf{x}_{1},\mathbf{x}_{2},\mathbf{x}_{3} \right)
& = & B_{N} \sum_{\mathcal{P}} (-1)^{\mathcal{P}} \mathcal{P} 
\sum_{M'm_{3}}
\sum_{m_{1}m_{2}} 
C_{J'M',\ j_{3}m_{3}}^{JM} 
C_{j_{1}m_{1},\ j_{2}m_{2}}^{J'M'}
\\
&& \times
\psi_{n_{1}\kappa_{1}m_{1}}\left(\mathbf{x}_{1}\right)
\psi_{n_{2}\kappa_{2}m_{2}}\left(\mathbf{x}_{2}\right)
\psi_{n_{3}\kappa_{3}m_{3}}\left(\mathbf{x}_{3}\right),
\label{eq:li_wf}
\end{eqnarray}
where $B_{N}$ is the normalization factor.
%
%
\\
\indent
%
%
Having constructed all the relevant parity conserving amplitudes, we 
now turn to evaluation of the parity violation in the resonance elastic 
electron scattering.
The dominant contribution to the PNC effect in the 
process of interest is provided by the nuclear spin-independent part of 
the weak interaction, which can be described by the following effective 
Hamiltonian~\cite{Khriplovich}
\begin{equation}
H_{\text{W}} = -\left(G_{\text{F}} / \sqrt{8} \right) Q_{\text{W}} 
\rho_{\text{N}}\left( r \right)\gamma_{5}.
\end{equation}
Here $Q_{\text{W}} \approx -N + Z\left(1 - 4
\sin^{2}\theta_{\text{W}}\right)$ is the weak charge of the nucleus, 
$\rho_{\text{N}}$ is the nuclear weak-charge density (normalized to 
unity), $G_{\text{F}}$ is the Fermi constant, and $\gamma_{5}$ is the 
Dirac matrix.
To account for the weak interaction we have to modify the wave 
functions:
\begin{eqnarray}
\left| \Psi_{d_{1}} \right\rangle \rightarrow 
\left| \Psi_{d_{1}} \right\rangle + 
\frac{
\left\langle \Psi_{d_{2}} \left| H_{\rm W} \right| \Psi_{d_{1}}
\right\rangle
}{E_{d_{1}} - E_{d_{2}}} \left| \Psi_{d_{2}} \right\rangle,
\label{eq:mod_d1}
\\
\left| \Psi_{d_{2}} \right\rangle \rightarrow 
\left| \Psi_{d_{2}} \right\rangle + 
\frac{
\left\langle \Psi_{d_{1}} \left| H_{\rm W} \right| \Psi_{d_{2}}
\right\rangle
}{E_{d_{2}} - E_{d_{1}}} \left| \Psi_{d_{1}} \right\rangle.
\label{eq:mod_d2}
\end{eqnarray}
To simplify the notations we define the admixing parameter 
${\rm i} \xi = \left\langle \Psi_{d_{1}} \left| H_{\rm W} \right| 
\Psi_{d_{2}} \right\rangle / \left(E_{d_{2}} - E_{d_{1}}\right)$.
Substituting Eqs.~(\ref{eq:mod_d1}) and~(\ref{eq:mod_d2}) into
Eq.~(\ref{eq:amplitude_2}) and keeping only the linear terms in $\xi$ 
one obtains the parity violating amplitude
\begin{eqnarray}
\nonumber
\tau_{\mu_{f}\mu_{i}}^{\rm PNC} & = & {\rm i} (2\pi)^{2} \varepsilon \xi
\sum_{M_{d}} 
\left(
      \left\langle \Psi_{f} \left|I\right| \Psi_{d_{2}} \right\rangle
      \left\langle \Psi_{d_{1}} \left|I\right| \Psi_{i} \right\rangle
    - \left\langle \Psi_{f} \left|I\right| \Psi_{d_{1}} \right\rangle
      \left\langle \Psi_{d_{2}} \left|I\right| \Psi_{i} \right\rangle
\right)
\\ && \times
  \left(
  \frac{1}{E_{i} - E_{d_{1}} + {\rm i} \Gamma_{d_{1}}/2}
- \frac{1}{E_{i} - E_{d_{2}} + {\rm i} \Gamma_{d_{2}}/2}
  \right).
\end{eqnarray}
Here we have utilized the fact that the weak interaction conserves the 
total momentum projection and, as a result, $M_{d}$ stands for 
$M_{d_{1}} = M_{d_{2}}$.
%
%
\\
\indent
%
%
One should point out that the nuclear spin-independent part of the weak 
interaction provides one more contribution to the PNC effect of the 
process studied.
This contribution is related to the scattering by the direct 
electron-nucleus weak interaction and can be expressed by the amplitude 
$(2 \pi)^{2} \varepsilon \left\langle \Psi_{f} \left| H_{\rm W} \right| 
\Psi_{i} \right\rangle$.
However, we omit this term since it is negligibly small in the framework
of the approximations considered.
Thus, the amplitude of the resonance elastic electron scattering 
can be written in the following form 
\begin{equation}
\tau_{\mu_{f}\mu_{i}} = \tau_{\mu_{f}\mu_{i}}^{\rm PC}
+ \tau_{\mu_{f}\mu_{i}}^{\rm PNC}
\end{equation}
with $\tau_{\mu_{f}\mu_{i}}^{\rm PC} = \tau_{\mu_{f}\mu_{i}}^{(0)} 
+ \tau_{\mu_{f}\mu_{i}}^{(1,\ \rm dir)}
+ \tau_{\mu_{f}\mu_{i}}^{(1,\ \rm exc)} + \tau_{\mu_{f}\mu_{i}}^{(2)}$
being the parity conserving contribution.
Examining the introduced amplitudes with respect to the spatial 
symmetry leads to the following rules
\begin{eqnarray}
&& \tau_{\mu\mu}^{\rm PC} = \tau_{-\mu-\mu}^{\rm PC},
\quad
\tau_{\mu-\mu}^{\rm PC} = -\tau_{-\mu\mu}^{\rm PC},
\label{eq:property_pc}
\\
&& \tau_{\mu\mu}^{\rm PNC} = -\tau_{-\mu-\mu}^{\rm PNC},
\quad
\tau_{\mu-\mu}^{\rm PNC} = \tau_{-\mu\mu}^{\rm PNC} = 0.
\label{eq:property_pnc}
\end{eqnarray}
%
%
%

\section{RESULTS AND DISCUSSION}
\label{sec:RD}
\begin{figure}[h!]
\caption{
The P-odd asymmetry of the resonance elastic electron scattering on 
He-like samarium ($Z = 62$) for two different scenarios.
In the first scenario (left graph) the polarization of the scattered 
electron is detected and in the second scenario the polarization is 
remained unobservable (right graph).
The solid and the dashed lines correspond to the cases of the incident 
electron energy being tuned in resonance with $\left[\left(1s 
2s\right)_{0} 7s\right]_{1/2}$ and $\left[\left(1s 2p_{1/2}\right)_{0} 
7s\right]_{1/2}$ states of the Li-like samarium, respectively.
}
\includegraphics[trim=0 0 0 0, clip, width = 1.0\textwidth]
{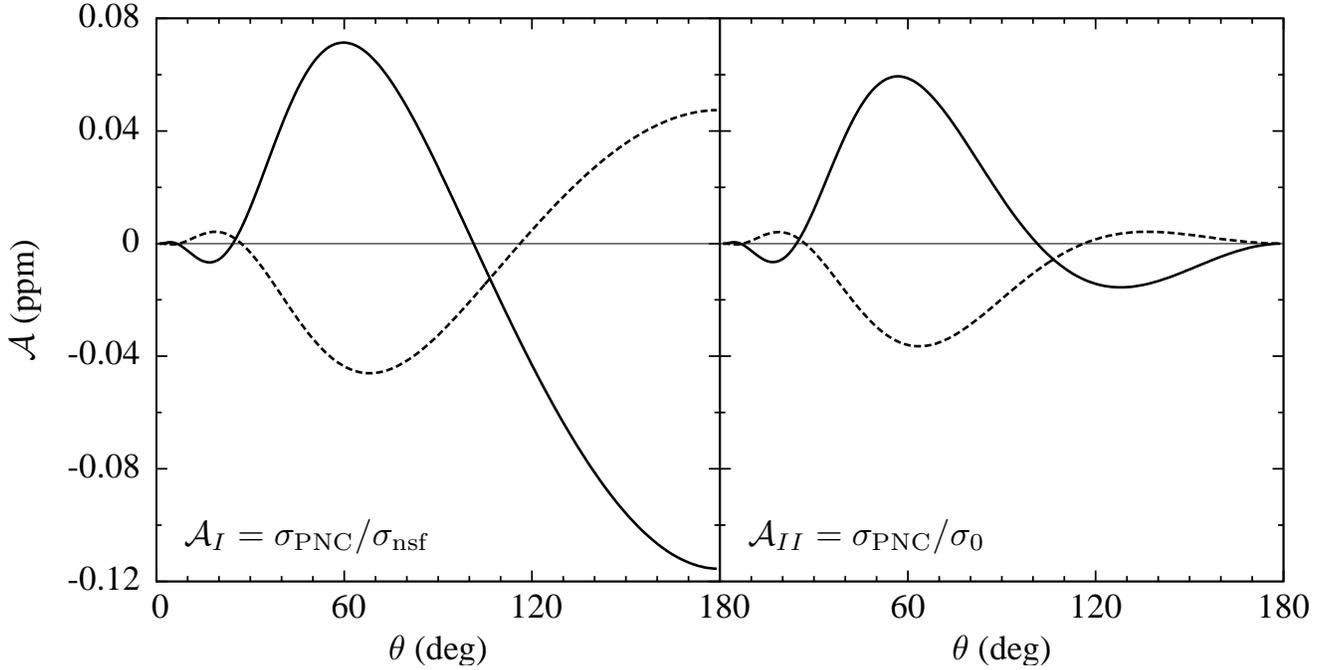} 
\label{fig:asymmetry_angles}
\end{figure}

\begin{figure}[h!]
\caption{
The asymmetry coefficients $\mathcal{A}_{I} = \sigma_{\rm PNC} / 
\sigma_{\rm nsf}$ (left graph) and $\mathcal{A}_{II} = \sigma_{\rm PNC}/ 
\sigma_{0}$ (right graph) of the resonance elastic electron scattering 
on He-like samarium ($Z = 62$).
The difference $E_{i} - E_{\left[\left(1s 2s\right)_{0}7s\right]_{1/2}}$
fixes the energy of the incoming electron.
The solid line corresponds to $\theta = 175^{\circ}$, the dashed and 
dotted lines are related to the cases of scattering at angles 
$110^{\circ}$ and $60^{\circ}$, respectively.
}
\includegraphics[trim=0 0 0 0, clip, width = 1.0\textwidth]
{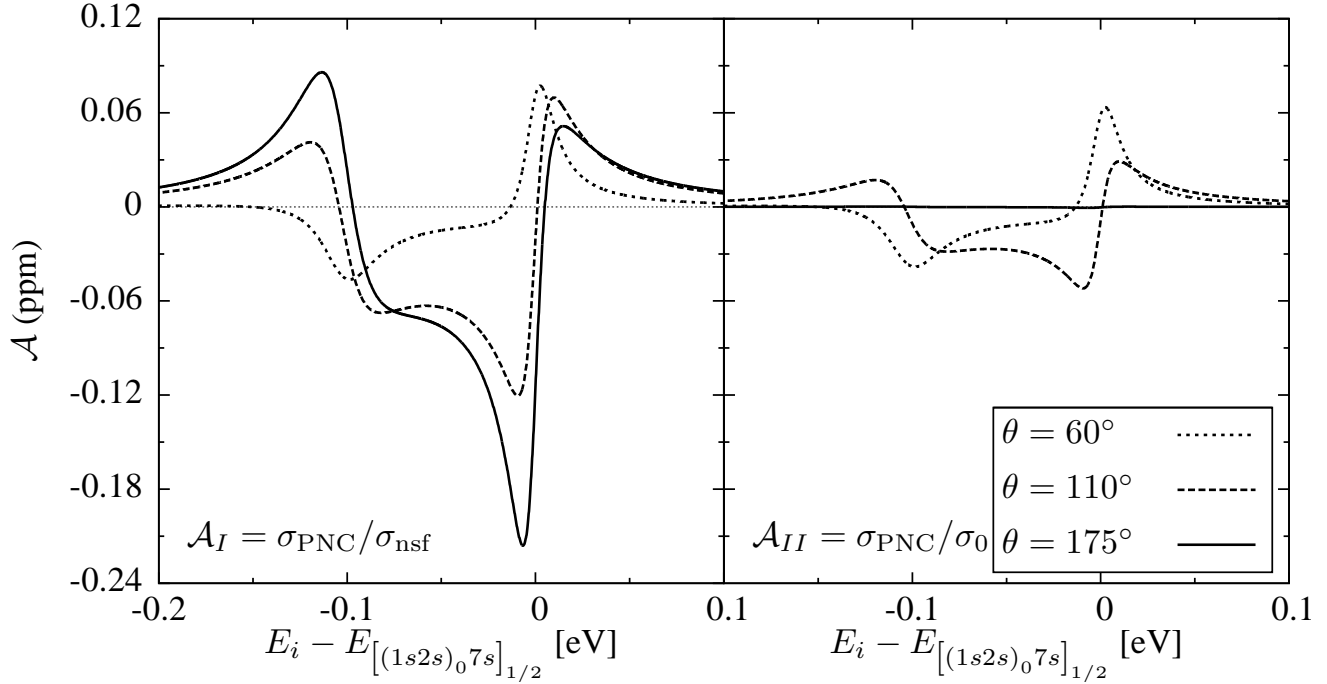} 
\label{fig:asymmetry_energy}
\end{figure}

%
%
In order to enhance the PNC effect in the elastic scattering of 
polarized electrons by He-like ions, being in the ground state, we 
assume that the energy of the incident electron is tuned in resonance 
with doubly-excited opposite-parity $d_{1} \equiv \levelp$ and $d_{2} 
\equiv \levels$ states of the corresponding Li-like ions.
The quasidegeneracy of these states was found for several $n, \kappa$ 
and $Z$ in Ref.~\cite{Zaytsev_PST156_014028:2013}.
%
%
\\
\indent
%
%
We study the influence of the parity violation on the differential cross 
section (DCS) $\sigma_{\mu_{f}\mu_{i}} \equiv d\sigma_{\mu_{f}\mu_{i}} / 
d\Omega = \left|\tau_{\mu_{f}\mu_{i}} \right|^{2}$ of the scattering 
process.
Let us introduce the non-spin-flip $\sigma_{\rm nsf} = 
\frac{1}{2} \left(\sigma_{1/2\ 1/2}\right. + 
\left.\sigma_{-1/2\ -1/2}\right)$ and 
the spin-flip $\sigma_{\rm sf} = \frac{1}{2} \left(\sigma_{1/2\ -1/2} + 
\sigma_{-1/2\ 1/2}\right)$ cross sections.
Then, the total DCS is $\sigma_{0} = \sigma_{\rm nsf} + \sigma_{\rm sf}$.
According to the rules~(\ref{eq:property_pnc}), the weak 
interaction modifies the cross section only in the case when the 
helicities of the incident and the outgoing electrons coincide 
($\mu_{i} = \mu_{f}$).
As a result, the presence of the PNC effect manifests in deviation of 
the P-odd contribution $\sigma_{\rm PNC} = \frac{1}{2}
\left(\sigma_{1/2\ 1/2} - \sigma_{-1/2\ -1/2}\right)$ to the cross 
section from zero.
In the present work we consider two scenarios.
In the first scenario, the polarization of the outgoing electron is 
assumed to be detected and only the non-spin-flip contribution to the 
cross section is considered.
In the second scenario the polarization remains unobserved and both 
$\sigma_{\rm nsf}$ and $\sigma_{\rm sf}$ are taken into account.
The luminosity of the first ($I$) and second ($II$) scenarios can be 
expressed as follows~\cite{Gribakin_PRA72_032109:2005, 
Maiorova_JPB42_205002:2009}
\begin{equation}
L_{I,\ II} =
\frac{\sigma_{I,\ II} + \sigma_{I,\ II}^{\rm (b)}}
{2\sigma_{\rm PNC}^{2} \eta^{2} T}.
\label{eq:luminosity}
\end{equation}
Here $\sigma_{I} = \sigma_{\rm nsf}$ while $\sigma_{II} = 
\sigma_{0}$, $\sigma_{I,\ II}^{\rm (b)}$ corresponds to the background 
signal, $T$ is the data collection time, and $\eta$ is the desired 
relative uncertainty of the PNC effect measurement.
In the present analysis we set $\sigma_{I,\ II}^{\rm (b)} = 0$, 
$T$ equals to two weeks, and $\eta = 1\%$.
%
%
\\
\indent
%
%
In Fig.~\ref{fig:asymmetry_angles}, the PNC asymmetry coefficients
$\mathcal{A}_{I} = \sigma_{\rm PNC} / \sigma_{\rm nsf}$ and
$\mathcal{A}_{II} = \sigma_{\rm PNC} / \sigma_{0}$ for the elastic 
electron scattering on He-like samarium ($Z = 62$) are displayed as
functions of the scattering angle $\theta$ in the case of resonance with
the $\left[\left(1s 2s\right)_{0} 7s\right]_{1/2}$ and 
$\left[\left(1s 2p_{1/2}\right)_{0} 7s\right]_{1/2}$ states.
Since these coefficients are directly related to the magnitude of the 
PNC effect, one can conclude that for the first scenario the parity 
violation is expected to become most significant at large scattering 
angles.
In the case when the polarization of the scattered electron is not 
detected (second scenario) the most promising situation occurs for 
$\theta \sim 60^{\circ}$, while at larger scattering angles a strong 
suppression of the P-odd asymmetry is observed.
This is due to the fact that at large scattering angles the dominant 
contribution to the DCS is provided by the P-even spin-flip amplitude, 
which does not interfere with the PNC amplitude according to
Eqs.~(\ref{eq:property_pnc}).
%
%
\\
\indent
%
%
In Fig.~\ref{fig:asymmetry_energy}, the parity violating asymmetry of 
the resonance electron scattering on He-like samarium ($Z = 62$) is 
depicted as a function of the incident electron energy for three 
different scattering angles (60, 110 and 175 degrees).
From this figure one can see that the peak magnitude of the P-odd 
asymmetry is expected for the energy of the scattering electron close to 
resonance which is provided by the $\left[\left(1s 2s\right)_{0}
7s\right]_{1/2}$ state.
Here it is worth to mention that the parameters, being related to the 
maximum magnitude of the parity violating asymmetry, may not provide the 
best value of the luminosity, and vice versa.
In order to find the optimal relation we propose the following 
procedure.
First, one should pick out scattering angles at which the P-odd 
asymmetry has the same order of magnitude as the maximal one.
Among them the optimal relation is provided by such an angle which 
corresponds to the minimum of the luminosity.
As an example, let us consider the scenario where the polarization of 
the outgoing electron is detected (first scenario).
For the case of the samarium ion (see Fig.~\ref{fig:asymmetry_energy}) 
the maximal value of $\mathcal{A}_{I}$ is expected for the 
scattering angle $175^{\circ}$ and equals $-2.2 \times 10^{-7}$, while
$L_{I}$ for these parameters is equal to $7.1 \times 10^{33}
$~cm$^{-2}$~s$^{-1}$.
The optimal relation between $\mathcal{A}_{I}$ and $L_{I}$ 
is expected for $\theta \sim 108^{\circ}$ where they take the values 
$-1.2\times10^{-7}$ and $6.5 \times 10^{31}$~cm$^{-2}$~s$^{-1}$, 
respectively.
%
%
\\
\indent
%
%
In Tables~\ref{tb:optimum_asym_lum_1scenario} and
\ref{tb:optimum_asym_lum_2scenario} we present the numerical results for 
the parameters $n$, $\kappa$ and $Z$ which seem to be most promising 
for measuring the PNC effect in the process of resonance elastic 
electron scattering on He-like ions.
It is assumed that the energy of the incident electron is tuned in 
vicinity of the resonance, being related to the $\levels$ state, to 
provide the peak value of the P-odd asymmetry.
In Table~\ref{tb:optimum_asym_lum_1scenario} we present the results for 
the case when the polarization of the scattered electron is measured 
(first scenario).
The results for the second scenario, where the polarization of the 
outgoing electron is not detected, are represented in 
Table~\ref{tb:optimum_asym_lum_2scenario}.
\begin{table}[h]
\begin{center}
\caption{
Cross section of the resonance elastic electron scattering on He-like 
ions for parameters $n$, $\kappa$ and $Z$ which seem to be most 
promising for measuring the PNC effect.
The energy of the incident electron is tuned in vicinity of resonance 
corresponding to the $\levels$ state.
It is assumed that the polarization of the scattered electron is 
detected.
The energy difference $\Delta E = E_{\levelp} - E_{\levels}$ is taken
from Ref.~\cite{Zaytsev_PST156_014028:2013}.
The scattering angle $\theta$ provides the optimal relation 
between the P-odd asymmetry $\mathcal{A}_{I} = \sigma_{\rm PNC} / 
\sigma_{\rm nsf}$ and the luminosity $L_{I}$, which is defined by 
Eq.~(\ref{eq:luminosity}).
$\sigma_{0}$ and $\sigma_{\rm nsf}$ are the total and non-spin-flip 
cross sections, respectively, and $\sigma_{\rm PNC}$ stands for the 
parity violating contribution to the cross section.
}
\label{tb:optimum_asym_lum_1scenario}
\end{center}
\begin{ruledtabular}
\begin{tabular}{cccccccccc}
$Z$ 
& $n\kappa$ 
& $\Delta E$ (eV) 
& $\varepsilon_{i}$ (keV) 
& $\theta$ (deg) 
& $\mathcal{A}_{I}$ 
& $L_{I}$ (cm$^{-2}$~s$^{-1}$) 
& $\sigma_{0}$ (b) 
& $\sigma_{\rm nsf}$ (b)
& $\sigma_{\rm PNC}$ (b)
\\ \hline
60 & $6s$ & -0.222(56) & 36.40 & 163 & $-1.0\times 10^{-7}$ 
& $2.7\times 10^{33}$ & $4.8\times 10^{3}$ & $1.5\times 10^{2}$ 
& $-1.5\times 10^{-5}$
\\
62 & $7s$ & -0.103(64) & 39.56 & 108 & $-1.2\times 10^{-7}$ 
& $6.5\times 10^{31}$ & $1.0\times 10^{4}$ & $4.7\times 10^{3}$ 
& $-5.5\times 10^{-4}$
\\
90 & $6s$ & 2.51(47) & 88.36 & 64 & $-3.3\times 10^{-8}$ 
& $1.7\times 10^{32}$ & $2.5\times 10^{4}$ & $2.2\times 10^{4}$ 
& $-7.2\times 10^{-4}$
\\
   & $7s$ & 1.75(47) & 89.22 & 57 & $3.8\times 10^{-8}$ 
& $9.3\times 10^{31}$ & $3.4\times 10^{4}$ & $3.0\times 10^{4}$ 
& $1.2\times 10^{-3}$
\\
92 & $5s$ & 2.97(28) & 91.43 & 66 & $-3.8\times 10^{-8}$ 
& $1.4\times 10^{32}$ & $2.3\times 10^{4}$ & $2.0\times 10^{4}$ 
& $-7.6\times 10^{-4}$
\\
   & $6s$ & -1.07(28) & 92.95 & 74 & $-1.0\times 10^{-7}$ 
& $3.1\times 10^{31}$ & $1.6\times 10^{4}$ & $1.3\times 10^{4}$ 
& $-1.3\times 10^{-3}$
\end{tabular}
\end{ruledtabular}
\end{table}
\begin{table}[h]
\begin{center}
\caption{
Cross section of the resonance elastic electron scattering on He-like 
ions for parameters $n$, $\kappa$ and $Z$ which seem to be most 
promising for measuring the PNC effect.
The energy of the incident electron is tuned in vicinity of resonance 
corresponding to the $\levels$ state.
It is assumed that the polarization of the scattered electron is not
detected.
The energy difference $\Delta E = E_{\levelp} - E_{\levels}$ is taken
from Ref.~\cite{Zaytsev_PST156_014028:2013}.
The scattering angle $\theta$ provides the optimal relation 
between the P-odd asymmetry $\mathcal{A}_{II} = \sigma_{\rm PNC} / 
\sigma_{0}$ and the luminosity $L_{II}$, which is defined by 
Eq.~(\ref{eq:luminosity}).
$\sigma_{0}$ and $\sigma_{\rm nsf}$ are the total and non-spin-flip 
cross sections, respectively, and $\sigma_{\rm PNC}$ stands for the 
parity violating contribution to the cross section.
}
\label{tb:optimum_asym_lum_2scenario}
\end{center}
\begin{ruledtabular}
\begin{tabular}{cccccccccc}
$Z$ 
& $n\kappa$ 
& $\Delta E$ (eV) 
& $\varepsilon_{i}$ (keV) 
& $\theta$ (deg) 
& $\mathcal{A}_{II}$ 
& $L_{II}$ (cm$^{-2}$~s$^{-1}$) 
& $\sigma_{0}$ (b) 
& $\sigma_{\rm nsf}$ (b)
& $\sigma_{\rm PNC}$ (b)
\\ \hline
60 & $6s$ & -0.222(56) & 36.40 & 43 & $ 2.2\times 10^{-8}$ 
& $4.0\times 10^{31}$ & $2.1\times 10^{5}$ & $1.9\times 10^{5}$ 
& $4.6\times 10^{-3}$
\\
62 & $7s$ & -0.103(64) & 39.56 & 45 & $ 4.9\times 10^{-8}$ 
& $1.0\times 10^{31}$ & $1.6\times 10^{5}$ & $1.5\times 10^{5}$ 
& $8.0\times 10^{-3}$
\\
90 & $6s$ & 2.51(47) & 88.36 & 59 & $-2.7\times 10^{-8}$ 
& $1.8\times 10^{32}$ & $3.2\times 10^{4}$ & $2.8\times 10^{4}$ 
& $-8.4\times 10^{-4}$
\\
   & $7s$ & 1.75(47) & 89.22 & 58 & $3.5\times 10^{-8}$ 
& $1.0\times 10^{32}$ & $3.2\times 10^{4}$ & $2.9\times 10^{4}$ 
& $1.1\times 10^{-3}$
\\
92 & $5s$ & 2.97(28) & 91.43 & 62 & $-3.2\times 10^{-8}$ 
& $1.5\times 10^{32}$ & $2.7\times 10^{4}$ & $2.4\times 10^{4}$ 
& $-8.5\times 10^{-4}$
\\
   & $5p_{1/2}$ & -0.511(27) & 91.44 & 46 & $2.2\times 10^{-8}$ 
& $1.2\times 10^{32}$ & $6.8\times 10^{4}$ & $6.3\times 10^{4}$ 
& $1.5\times 10^{-3}$
\end{tabular}
\end{ruledtabular}
\end{table}
%
%
\\
\indent
%
%
From Table~\ref{tb:optimum_asym_lum_1scenario} one can see that 
for the first scenario the PNC effect is expected to be most 
pronounced for scattering on the samarium ($Z = 62$) ion at the energy 
of the incident electron tuned in vicinity of resonance corresponding to 
the $\left[\left(1s 2s\right)_{0}7s\right]_{1/2}$ state.
In this case, the optimal values of the asymmetry and the luminosity 
equal to $-1.2 \times 10^{-7}$ and $6.5 \times 10^{31}$
cm$^{-2}$~s$^{-1}$, respectively, and are achieved for the scattering 
angle $\sim 108^{\circ}$.
This system seems also to be most preferable for the second scenario 
(Table~\ref{tb:optimum_asym_lum_2scenario}), where the polarization of 
the scattered electron is not detected.
In this scenario the optimal values $\mathcal{A}_{II} = 4.9 \times 
10^{-8}$ and $L_{II} = 1.0 \times 10^{31}$ cm$^{-2}$~s$^{-1}$ are 
obtained at $\theta \sim 45^{\circ}$.
From these tables one can conclude that the observation of the outgoing
electron polarization does not allow to increase significantly the PNC 
effect.
%
%
\\
\indent
%
%
The requirement of the electron production with high and controllable 
degree of spin polarization and accurate energy tuning makes it 
presently impossible to investigate the P-odd effects in the process of 
interest.
Perhaps, some of the difficulties can be avoided by studying inelastic 
electron scattering, where one could get rid of the dominant zero-order 
(in $1/Z$) contribution to the PC amplitude, thus reducing the 
suppression of the PNC effect.
One may also think, that the corresponding investigations with other 
heavy few-electron ions can lead to a bigger effect.
We expect that the calculations performed in the present paper can serve 
as a proper basis for further study in these directions.
%

\section{CONCLUSION}
\label{sec:CON}
%
%
In the present work the PNC effect has been studied in the elastic 
scattering of polarized electrons by heavy He-like ions, being initially 
in the ground state.
In order to enhance the parity violation effect, the energy of the 
incident electron has been chosen to provide a resonance with one of the 
quasidegenerate doubly-excited $\levels$ and $\levelp$ states of the 
corresponding Li-like ion.
We have considered two different scenarios.
In the first scenario we assume that the polarization of the scattered 
electron was measured.
In the second one the polarization was supposed to be unobservable.
It has been found that for both variants the PNC effect occurs to be  
most pronounced for scattering on samarium ion at the energy of the 
incident electron tuned in vicinity of resonance, 
which is related to the $\left[\left(1s 2s\right)_{0} 7s\right]_{1/2}$ 
state.
In the case of the first scenario the peak value of the PNC asymmetry 
equals to $-1.2 \times 10^{-7}$ at $\theta \sim 108^{\circ}$, while in the
second scenario the P-odd asymmetry is $4.9 \times 10^{-8}$ for the 
scattering angle $\theta \sim 45^{\circ}$.
These values are too small to make possible performing the corresponding
experiment.
We think, however, that the calculations presented can be considered as 
the first necessary step towards investigations of the PNC effect with 
electron scattering by heavy ions.
%

\section{ACKNOWLEDGEMENTS}
%
Fruitful discussions with D.~A.~Telnov are gratefully acknowledged.        
This work was supported by the grant of the President of the Russian 
federation (Grant No. MK-1676.2014.2), by RFBR (Grant No. 13-02-00630),
by SPbSU (Grant No. 11.38.269.2014), and by SAEC Rosatom.
Additional support was received from DFG, GSI, and DAAD. 
The work of V.A.Z. was supported by the German-Russian Interdisciplinary 
Science Center (G-RISC).
V.A.Z. and A.V.M. acknowledges financial support by the \textquotedblleft 
Dynasty\textquotedblright~foundation.
S.T. acknowledges support by the German Research Foundation (DFG) within 
the Emmy Noether program under Contract No.~TA~740~1-1.


%


\begin{thebibliography}{99}
\bibitem{Khriplovich}
I.~B.~Khriplovich,
\textit{Parity Nonconservation in Atomic Phenomena} (Gordon and Breach, 
London, 1991).
\bibitem{Khriplovich_PST112_52:2004}
I.~B.~Khriplovich,
Phys. Scr. \textbf{T112}, 52 (2004).
\bibitem{Ginges_PR397_63:2004}
J.~S.~M.~Ginges and   V.~V.~Flambaum,
Phys. Rep. {\bf 397}, 63 (2004).
\bibitem{Wood_S275_1759:1997}
C.~S.~Wood, S.~C.~Bennett, D.~Cho, B.~P.~Masterson, J.~L.~Roberts, 
C.~E.~Tanner, and C.~E.~Wieman,
Science \textbf{275}, 1759 (1997).
\bibitem{Bennett_PRL82_2484:1999}
S.~C.~Bennett and C.~E.~Wieman,
Phys. Rev. Lett. \textbf{82}, 2484 (1999);
Phys. Rev. Lett. \textbf{83}, 889 (1999).
\bibitem{Shabaev_PRL94_213002:2005}
V.~M.~Shabaev, K.~Pachucki, I.~I.~Tupitsyn, and V.~A.~Yerokhin,
Phys. Rev. Lett. \textbf{94}, 213002 (2005);
V.~M.~Shabaev, I.~I.~Tupitsyn, K.~Pachucki, G.~Plunien, and V.~A.~Yerokhin,
Phys. Rev. A \textbf{72}, 062105 (2005).
\bibitem{Porsev_PRL102_181601:2009}
S.~G.~Porsev, K.~Beloy, and A.~Derevianko,
Phys. Rev. Lett. \textbf{102}, 181601 (2009).
\bibitem{Dzuba_PRL109_203003:2012}
V.~A.~Dzuba, J.~C.~Berengut, V.~V.~Flambaum, and B.~Roberts,
Phys. Rev. Lett. {\bf 109}, 203003 (2012).
\bibitem{Gorshkov_JL19_394:1974}
V.~G.~Gorshkov and L.~N.~Labzowsky,
Zh.~Eksp.~Teor.~Fiz.~Pis'ma \textbf{19}, 768 (1974)
[JETP Lett. \textbf{19}, 394 (1974)];
Zh.~Eksp.~Teor.~Fiz. \textbf{69}, 1141 (1975)
[Sov. Phys. JETP \textbf{42}, 581 (1975)].
\bibitem{Pindzola_PRA47_4856:1993}
M.~S.~Pindzola, Phys. Rev. A \textbf{47}, 4856 (1993).
\bibitem{Gribakin_PRA72_032109:2005}
G.~F.~Gribakin, F.~J.~Currell, M.~G.~Kozlov, and A.~I.~Mikhailov,
Phys. Rev. A \textbf{72}, 032109 (2005); Phys. Rev. A \textbf{80},
049901(E) (2009); arXiv:physics/0504129 (2005).
\bibitem{Zaytsev_PRA89_032703:2014}
V.~A.~Zaytsev, A.~V.~Maiorova, V.~M.~Shabaev, A.~V.~Volotka, S.~Tashenov, 
G.~Plunien, and Th.~St\"ohlker,
Phys. Rev. A {\bf 89}, 032703 (2014).
\bibitem{Maiorova_JPB42_205002:2009}
A.~V.~Maiorova, O.~I.~Pavlova, V.~M.~Shabaev, C.~Kozhuharov, G.~Plunien, 
and Th.~St\"ohlker,
J. Phys. B: At. Mol. Opt. Phys. {\bf 42}, 205002 (2009).
\bibitem{Maiorova_JPB44_225003:2011}
A.~V.~Maiorova, V.~M.~Shabaev, A.~V.~Volotka, V.~A.~Zaytsev,
G.~Plunien, and Th.~St\"ohlker, J. Phys. B: At. Mol. Opt. Phys.
{\bf 44}, 225003 (2011).
\bibitem{Gunst_PRA87_032714:2013}
J.~Gunst, A.~Surzhykov, A.~Artemyev, S.~Fritzsche, S.~Tashenov, 
A.~V.~Maiorova, V.~M.~Shabaev, and Th.~St\"ohlker,
Phys. Rev. A {\bf 87}, 032714 (2013).
\bibitem{Shabaev_PRA81_052102:2010}
V.~M.~Shabaev, A.~V.~Volotka, C.~Kozhuharov, G.~Plunien, and Th.~St\"ohlker,
Phys. Rev. A {\bf 81}, 052102 (2010).
\bibitem{Surzhykov_PST156_014027:2013}
A.~Surzhykov, A.~V.~Maiorova, V.~M.~Shabaev, Th.~St\"ohlker, and 
S.~Fritzsche
Phys. Scr. {\bf T156}, 014027 (2013).
\bibitem{Zaytsev_PST156_014028:2013}
V.~A.~Zaytsev, A.~V.~Maiorova, V.~M.~Shabaev, A.~V.~Volotka, and 
G.~Plunien,
Phys. Scr. {\bf T156}, 014028 (2013).
\bibitem{Akhiezer:1965}
A.~I.~Akhiezer and V.~B.~Berestetsky,
\textit{Quantum Electrodynamics} 
(Wiley, New York, 1965).
\bibitem{Berestetsky:2006}
V.~B.~Berestetsky, E.~M.~Lifshitz, and L.~P.~Pitaevskii,
\textit{Quantum Electrodynamics} 
(Butterworth-Heinemann, Oxford, 2006).
%
\bibitem{Mott_PRS135_429:1932}
N.~F.~Mott,
Proc. R. Soc. (London) {\bf 135}, 429 (1932).
%
\bibitem{Dogget_PR103_1597:1956}
J.~A.~Dogget and L.~V.~Spencer,
Phys. Rev. {\bf 103}, 1597 (1956).
%
\bibitem{Sherman_PR103_1601:1956}
N.~Sherman,
Phys. Rev. {\bf 103}, 1601 (1956).
%
\bibitem{Gluckstern_JMP5_1594:1964}
R.~L.~Gluckstern and S.~R.~Lin,
J. Math. Phys. {\bf 5}, 1594 (1964).
%
\bibitem{Maiorova_JPB43_205006:2010}
A.~V.~Maiorova, D.~A.~Telnov, V.~M.~Shabaev, V.~A.~Zaytsev, G.~Plunien,
and Th.~St\"ohlker,
J. Phys. B: At. Mol. Opt. Phys. {\bf 43}, 205006 (2010).
%
\bibitem{Shabaev_PRA50_4521:1994}
V.~M.~Shabaev,
Phys. Rev. A {\bf 50}, 4521 (1994).
%
\bibitem{Shabaev_PR356_119:2002}
V.~M.~Shabaev,
Phys. Rep. \textbf{356}, 119 (2002).
%
\bibitem{Rose:1957}
M.~E.~Rose,
\textit{Elementary Theory of Angular Momentum} 
(Wiley, New York, 1957).
%
\bibitem{Eichler_PR439_1:2007}
J.~Eichler and Th.~St\"ohlker,
Phys. Rep. {\bf 439}, 1 (2007).
%
\bibitem{Varshalovich:1988}
D.~A.~Varshalovich, A.~N.~Moskalev, and V.~K.~Khersonskii,
\textit{Quantum Theory of Angular Momentum} 
(World Scientific, Singapore, 1988).
%
\end{thebibliography}
\end {document}